\documentclass[11pt]{article}
\usepackage{graphicx}
\usepackage{amssymb,amsmath}

\newcommand{\jprlBase}       {Phys.\ Rev.\ Lett.\xspace}
\newcommand{\jprl}      [1]  {\jprlBase\ {\bf #1}}

\newcommand{\jprBase}        {Phys.\ Rev.\xspace}
\newcommand{\jprd}      [1]  {\jprBase\ D~{\bf #1}}

\newcommand{\progtp}    [1]  {{Prog.\ Theor.\ Phys.\ {\bf #1}}}
\newcommand{\jplBase}        {Phys.\ Lett.\xspace}
\newcommand{\plb}       [1]  {\jplBase\ B~{\bf #1}}
\newcommand{\npBase}         {Nucl.\ Phys.\xspace}
\newcommand{\npb}       [1]  {\npBase\ B~{\bf #1}}
\newcommand{\nimBaseA}       {Nucl.\ Instr.\ Meth.\xspace}
\newcommand{\nima}      [1]  {\nimBaseA~A~{\bf #1}}

\newcommand{\npbps}     [1]  {{Nucl.\ Phys.\ B~Proc.\ Suppl.\ {\bf #1}}}

\newcommand{\BABARPubYear}    {06}

\newcommand{\BABARProcNumber} {90}
\newcommand{\SLACPubNumber} {12153}

\input definitions

\long\def\inst#1{\par\nobreak\kern 4pt\nobreak
    {\it #1}\par\vskip 10pt plus 3pt minus 3pt}

\begin{document}
{\pagestyle{empty}

\begin{flushright}
SLAC-PUB-\SLACPubNumber \\
\babar-PROC-\BABARPubYear/\BABARProcNumber \\
October, 2006 \\
\end{flushright}

\par\vskip 4cm

\begin{center}
\Large \bf Measurements of the CKM angle $\beta$ in charmless loop-dominated
  $B$ meson decays at \babar
\end{center}
\bigskip

\begin{center}
\large
Alfio Lazzaro \\
Dipartimento di Fisica and INFN, Universit\`a degli Studi di Milano \\
via Celoria 16, I-20133 Milano, Italy \\
(representing the \babar\ Collaboration)
\end{center}
\bigskip \bigskip

\begin{center}
\large \bf Abstract
\end{center}
We report on preliminary measurements of time-dependent \CP-violation
asymmetries in charmless neutral $B$ meson decays to \kpkmkz
(including resonant decays \phikz\ and \fzkz),
\etaprkz, \pizkzs, \kzskzskzs,
\kzskzs, \rhozkzs, \omegakzs. The results are obtained from a
data sample of up to $347$ million \BB\ pairs produced by \epem\ annihilation
at the \FourS\ resonance collected
with the \babar\  detector at the \pep2 asymmetric-energy $B$-meson
Factory at SLAC.

\vfill
\begin{center}
Contributed to the Proceedings of the 33$^{th}$ International 
Conference on High Energy Physics, \\
7/26/2006---8/2/2006, Moscow, Russian Federation. 
\end{center}

\vspace{1.0cm}
\begin{center}
{\em Stanford Linear Accelerator Center, Stanford University, 
Stanford, CA 94309} \\ \vspace{0.1cm}\hrule\vspace{0.1cm}
Work supported in part by Department of Energy contract DE-AC02-76SF00515.
\end{center}

\section{Introduction}

Measurements of time-dependent \CP\ asymmetries in $B^0$ meson decays
through a dominant Cabibbo-Kobayashi-Maskawa (CKM) favored $b \rightarrow c
\bar{c} s$ amplitude~\cite{sin2betaD} have provided a crucial test of the
mechanism of \CP\ violation in the Standard Model (SM)~\cite{SM}.
For such decays the interference between this amplitude and 
\BzBzb\ mixing is dominated by the single phase $\beta = \arg{(-V_{cd}
V^*_{cb}/V_{td} V^*_{tb})}$ of the CKM mixing matrix.
The \babar\ measurement of $\beta$ for these modes is $\stwob=0.710 \pm
0.039$~\cite{s2bBabar}.

In the SM, decays of $B^0$ mesons to charmless hadronic final states, such as
\phikz, \fzkz, \kpkmkz, \etaprkz, \pizkzs, \kzskzskzs, \rhozkzs, \omegakzs,
proceed 
mostly via a single loop (penguin)
amplitude with the same weak phase as the $b \to c \bar{c} s$
transition~\cite{Penguin}. In these modes, assuming the penguin
dominance of \btos\ transition and neglecting CKM-suppressed amplitudes, the
time-dependent 
\CP-violation parameter 
$S$ (defined in Eq.~\ref{eq:FCPdef} below) is expected to be \stwob.
However, CKM-suppressed amplitudes and the color-suppressed
tree-level diagram, present in not \mbox{pure-penguin} modes
(like \fzkz, \kpkmkz, \etaprkz, \pizkzs, \rhozkzs,
\omegakzs), 
introduce additional 
weak phases whose contribution may not be
negligible~\cite{Gross,Gronau,BN,london}. As a 
consequence, only an effective $S = \stwob_{\rm eff}$
is determined. The deviation $\Delta S=S - \stwob$
has been estimated in
several theoretical approaches: QCD factorization (QCDF)~\cite{BN,BN2}, QCDF
with modeled 
rescattering~\cite{Cheng}, Soft Collinear Effective Theory (SCET)~\cite{Zupan},
and SU(3) symmetry~\cite{Gross,Gronau,Jonat}. The estimates are
channel-dependent. QCDF and SCET models estimate
$\Delta S$ to be positive in the most of modes. SU(3)
symmetry provides unsigned bounds
of the order $|\Delta S | \lesssim 0.05$ in the best case. 

Due to the large virtual mass scales occurring in the penguin loops,
the possible presence of additional diagrams with new heavy particles in
the loop and new
\CP-violating phases may contribute to the decay amplitudes.
In this case the measurements of significantly larger $\Delta S$ are a
sensitive probe for physics beyond the SM~\cite{london}. Due to the different
non-perturbative strong-interaction properties 
of the various penguin decays,  the effect of New
Physics (NP) is expected to be channel-dependent.

In the SM, the decay \btokzskzs\ is expected to be dominated by the penguin
\btod\ transition, and is potentially sensitive to the presence of NP in a way
analogous to \btos\ decays~\cite{LondonQuinn}. Neglecting CKM-suppressed
amplitudes, the time-dependent \CP-violating asymmetry parameters in this mode
are expected to vanish (i.e. $S=0$), while contributions from lighter quarks
or supersymmetric particles could induce observable asymmetries~\cite{Giri}.

In this summary we report preliminary measurements of \CP-violating
asymmetries in all of the above mentioned loop-dominated decays.
The data sample used consists of $347$
 million
\BB\ pairs ($227$
million for \rhozkzs), recorded at the $\Upsilon (4S)$
resonance (center-of-mass energy 
$\sqrt{s}=10.58\ \gev$). The data were collected with the
\babar\  detector~\cite{BABARNIM} at the PEP-II asymmetric-energy
\epem\ collider. 
Detailed description for each analysis presented here are
given in Refs.~\cite{details1} and~\cite{details2}. 
In Ref.~\cite{BABARNIM} we describe the silicon vertex tracker
(SVT) and drift chamber used for track and vertex reconstruction, the
Cherenkov detector (DIRC), the electromagnetic calorimeter (EMC),
and the instrumented flux return (IFR).


We reconstruct a \Bz\ decaying into the
\CP\ eigenstates \kpkmkz, \etaprkz, \pizkzs,
\kzskzskzs, 
\kzskzs, \rhozkzs, \omegakzs\ ($B_{\CP}$).  From the
remaining particles in the event we also reconstruct the decay
vertex of the other $B$ meson ($B_{\rm tag}$) and identify its flavor.
The difference $\deltat \equiv t_{\CP} - t_{\rm tag}$ of the proper decay
times $\tcp$ and $\ttag$ of the \CP\ and tag $B$ mesons, respectively, is
obtained from the measured distance between the $B_{\CP}$ and $B_{\rm
tag}$ decay vertices and from the boost ($\beta \gamma =0.56$) of the
\epem\ system. Due to the \KS\ lifetime,
the \dt\ for the modes \pizkzs,
\kzskzskzs, \kzskzs is obtained reliably by
exploiting the 
knowledge of the average interaction 
point and from a global constrained fit to the entire $\Y4S\to\Bz\Bzb$ decay
tree, including the constraint 
from the lifetime of the $\Bz$ meson. 

The \deltat\ distribution
is given by: 
\begin{equation}
  F(\dt) =
        \frac{e^{-\left|\deltat\right|/\tau}}{4\tau} [1 \mp\Delta w \pm
   (1-2w) (-\eta_f S\sin(\deltamd\deltat) - 
   C\cos(\deltamd\deltat))], 
\label{eq:FCPdef}
\end{equation}
where $\eta_f$ is the \CP\ eigenvalue of the final
state $f$, the upper (lower) sign denotes a decay accompanied by a \Bz (\Bzb)
tag, 
$\tau$ is the mean $\Bz$ lifetime, $\deltamd$ is the $\BzBzb$ mixing
frequency, 
and the mistag parameters $w$ and $\Delta w$ are the average and
difference, respectively, of the probabilities that a true $\Bz$\ is
incorrectly tagged as a $\Bzb$\ or vice versa.  The tagging algorithm
has six mutually exclusive tagging categories~\cite{s2bBabar}.  A non-zero
value of the 
parameter $C$ would indicate direct \CP\ violation.

\section{Analysis Method}

Considering the $\KS$ candidates reconstructed in $\pi^+\pi^-$,
we reconstruct the signal candidates combining 
 $\KS$  and
$\Kp\Km$, $\etapr$, $\piz$, $\KS\KS$,
$\rhoz$, $\omega$, or another $\KS$ candidate. 
We also reconstruct the $\KS$ candidates in 
$\piz\piz$, and they are combined with $\Kp\Km$, $\etapr$, or $\KS\KS$ (both
in $\pi^+\pi^-$). Finally, we reconstruct the modes $\Bz \to \Kp\Km\KL$ and 
$\Bz \to \etapr\KL$, where a
$\KL$ candidate is identified either as an unassociated
cluster of energy in the EMC or as a cluster of hits in
the IFR. The $\etapr$ candidates are reconstructed
in $\rhoz\gamma$ and in $\eta \pip \pim$, with $\eta$ candidates in
$\pip\pim\piz$ (not considered in the signal candidates with $\KS \to
\piz\piz$  and $\KL$) and in $\gamma\gamma$.  


We use the informations from the tracking system, the EMC and the DIRC 
to identify pions and kaons in the final state. 
Two kinematic variables are used to 
discriminate between signal decays and combinatorial
background. The first is $\Delta E$, the difference between the center-of-mass
(CM) energy of the $B$~candidate and the CM beam energy. 
The second is the beam-energy-substituted mass $\mes \equiv \sqrt{(s/2 +
\pvec_0\cdot\pvec_B)^2/E_0^2 - \pvec_B^2}$, where the $B$
candidate momentum $\pvec_B$ and the four-momentum of the initial \FourS\
state $(E_0,\pvec_0)$ are defined in the laboratory frame.
In the $\piz\KS$ and $\KS\KS\KS$ analyses, these variables are replaced
by the invariant mass of the reconstructed $B$ meson, $m_B$, 
and the missing mass $m_{\rm miss}=|q_{\epem}-\hat q_{B}|$, where
$q_{\epem}$ is the four-momentum of the $\epem$ system
and $\hat q_B$ is the four-momentum of the $B$ candidate
after applying a \Bz-mass constraint.

Background events arise primarily from random combinations of particles in
continuum $\epem\to\qqbar$ events ($q=u,d,s,c$).  We reduce these with
requirements on shape-event variables, like the angle \thetaT\ between
the thrust axis of the $B$ 
candidate in the \FourS\ frame and that of the rest of the charged
tracks and neutral calorimeter clusters in the event.
In the fit we discriminate further against \qqbar\ background with a
Fisher discriminant \xf\ or a neural network (NN) which combines several
variables that 
characterize the production dynamics and energy flow in the
event~\cite{AngMom}. We study the background from other $B$ decays using Monte
Carlo (MC) simulated events. We take care of this background adding specific
components in the fit. 

We obtain the \CP-violation parameters and signal yields for each
mode from extended maximum likelihood fits with the input observables $\Delta
E$, \mes, \xf\ or NN, \deltat\ (all modes) as well as the resonance mass and
decay angle (\rhozkzs\ and \omegakzs). In the case of modes with \KL\ the
$B$ mass kinematic 
constraint is necessary to determine the $\KL$ momentum so that
$\mes$ cannot be exploited in the fit.
In the fits, the likelihood for a given event is
the sum of the signal, continuum and the $B$-background
likelihoods, weighed by their respective event yields.

In the \kpkmkz\ analysis, we use an angular moment analysis to extract
strengths of the partial waves in $\Kp \Km$ mass bins. In this approach we
rely only on the assumption that the two lowest partial waves ($S$- and
$P$-wave) are present,
but make no other assumption on the decay model. Furthermore, for this mode 
we develop a novel technique based on a
time-dependent Dalitz plot analysis, to take into account the variation of
\CP\ content and interference naturally in the fit. We use an isobar model
where we include the resonances $f_0(980)$, $\phi(1020)$, $X_0(1550)$, and
$\chi_{c0}$. In addition to resonant decays, we include non-resonant
amplitudes. We extract $\beta_{\rm eff}$ and $A_{\CP} =
-C$ from the asymmetries in amplitudes and phases between \Bz\ and 
\Bzb\ decays across the Dalitz plot. 

\section{Results}

The fit results for the \CP parameters are given
in Table~\ref{tab:results}. In \kpkmkz\ analysis, we find that the
trigonometric 
reflection at $\pi/2 - \beta_{\rm eff}$ is disfavored at $4.6\sigma$, which is
the first such measurement in penguin decays.
From the angular moment analysis we 
find the $P$-wave fraction to be $0.29\pm0.03$ averaged over the Dalitz plot,
and 
$0.89\pm0.01$ over the
$\phi$ resonance region ($1.0045 < m_{\Kp\Km} < 1.0345$~\gevcc).
We perform also a fit to 
low-$\Kp\Km$ mass region ($m_{\Kp\Km} <1.1~\gevcc$)
in order to measure \CP-asymmetry parameters for 
$\phi \Kz$ and $f_0(980) \Kz$
components. This is the first measurement of the \CP\ parameters in 
$f_0(980) \Kz$ with $f_0(980)\to \Kp\Km$. 

\begin{table}[t]
\caption{Preliminary fit results for each penguin mode. 
  The second column gives
  $\beta_{\rm eff}$ or $S$, the third the result for the $A_{\CP}$ or $C$. The
  first errors given are 
  statistical and the second systematic. See text for the description of
  $\phi \Kz$ and $f_0(980)\Kz$ results.
  \label{tab:results}}
{
\setlength{\tabcolsep}{0.25pc}
\begin{center}
\begin{tabular}{@{}ccc@{}}
\hline \hline
Mode & $\beta_{\rm eff}$ & $A_{\CP}$ \\ \hline
\multicolumn{3}{l}{\hspace{-0.2pc}\boldmath{\kpkmkz}} \\
& $0.361\PM0.079\PM 0.037$ & $-0.034\PM0.079\PM 0.025$\\
\multicolumn{3}{l}{~~$\phi \Kz$} \\
& $0.06\PM0.16\PM 0.05$&  $-0.18\PM 0.20 \PM 0.10$ \\
\multicolumn{3}{l}{~~$f_0(980)\Kz$} \\
& $0.18\PM0.19\PM 0.04$&  $\msp0.45\PM 0.28 \PM 0.10$ \\
\hline
Mode & $S$ & $C$ \\ \hline
\multicolumn{3}{l}{\hspace{-0.2pc}\boldmath{\etaprkz}} \\
& $\msp0.55\PM0.11 \PM 0.02$ &  $-0.15\PM 0.07 \PM 0.03$\\
\multicolumn{3}{l}{\hspace{-0.2pc}\boldmath{\pizkzs}} \\
& $\msp0.33\PM0.26 \PM 0.04$ & $\msp0.20\PM0.16 \PM 0.03$\\
\multicolumn{3}{l}{\hspace{-0.2pc}\boldmath{\kzskzskzs}} \\
& $\msp0.66\PM0.26 \PM 0.08$ &  $-0.14\PM 0.22 \PM 0.05$\\
\multicolumn{3}{l}{\hspace{-0.2pc}\boldmath{\rhozkzs}} \\
& $\msp0.17\PM0.52\PM0.26$ &  $\msp0.64\PM0.41\PM0.25$\\
\multicolumn{3}{l}{\hspace{-0.2pc}\boldmath{\omegakzs}} \\
& $\msp0.62^{+0.25}_{-0.30}\PM0.02$ &$-0.43^{+0.25}_{-0.23} \PM 0.03$\\
\multicolumn{3}{l}{\hspace{-0.2pc}\boldmath{\kzskzs}} \\
& $-1.28^{+0.80\;+0.11}_{-0.73\;-0.16}$ & $-0.40\PM0.41\PM 0.06$\\
\hline
\hline
\end{tabular}
\end{center}
}
\end{table}

All measurements reported here are statistics limited. The main sources of
systematic errors are the Dalitz plot model (\kpkmkz), the \CP content of \BB\
background, the reference
shape modeling, the interference with other resonances ($\rhoz\KS$), and
the SVT alignment (\pizkzs, \kzskzskzs, \kzskzs). 

The individual $\stwob_{\rm eff}$ results are in agreement with the
charmonium value. The $\stwob_{\rm eff}$ for \etaprkz\ is 
inconsistent with zero by $4.9$ standard deviations.
The $S$ value for \kzskzs\ is consistent with zero.
None of the modes studied exhibits non-zero direct \CP violation.

\section{Conclusion}

We have presented preliminary results on mixing-induced and direct \CP
violation for loop-dominated $B$ decays. We have studied several
modes, reconstructing several 
sub-decays for each mode in order to increase the statistics.
We have improved the analyses using better techniche, like time-dependent
Dalitz plot analysis. 

All individual penguin modes are in agreement with the SM expectation.
However, the $\stwob_{\rm eff}$ measurements are lower than the charmonium
value so  
that their average appears to be rather low. With the increase of
statistics, we are very close to measure the \CP\ violation in the penguin
modes with \btos\ transition.

\section*{Acknowledgments}

I thank my \babar\ colleagues for their support, in particular Riccardo
Faccini, Denis Dujmic, and Fernando Palombo. I would like to thank
also my friend 
Riccardo Cirillo for his friendly support during the conference.

\end{document}